\documentclass[sigconf]{acmart}

\usepackage{booktabs} 
\usepackage{lmodern}
\usepackage{array}
\usepackage{rotfloat}
\usepackage{multirow}
\usepackage{graphicx}
\usepackage{colortbl}
\usepackage{a4wide}
\usepackage{float}
\usepackage{stfloats}
\usepackage{pdflscape}
\usepackage{float}
\usepackage{flushend}
\floatstyle{boxed} 

\newboolean{showcomments}
\setboolean{showcomments}{true} 
\ifthenelse{\boolean{showcomments}}
{\newcommand{\nb}[2]{
  \fcolorbox{black}{yellow}{\bfseries\sffamily\scriptsize#1}
  {\sf\small$\blacktriangleright$\textit{#2}$\blacktriangleleft$}
 }
 
}
{\newcommand{\nb}[2]{}
 
}

\let\textquotedbl="
\setcopyright{rightsretained}

\acmDOI{xx.xxx/xxx_x}

\acmISBN{xxx-x-xxxx-xxxx-x/xx/xx}

\acmConference[CONF]{Sample Conference}{November, 2018}{} 
\acmYear{2018}
\copyrightyear{2018}

\acmArticle{4}
\acmPrice{15.00}

\begin{document}
\title[Blockchain and Cryptocurrency: A Comparative Framework]{Blockchain and Cryptocurrency: A comparative framework of the main Architectural Drivers}

\author{Martin Garriga}
\affiliation{%
  \institution{CONICET, National Council for Scientific and Technical Research}
  \streetaddress{Buenos Aires 1400}
  \state{Neuquen} 
  \postcode{8300}
  \country{Argentina}
}
\email{martin.garriga@fi.uncoma.edu.ar}

\author{Maxmiliano Arias}
\affiliation{%
  \institution{Fidtech}
  \streetaddress{Independencia 596}
  \state{Neuquen} 
  \postcode{8300}
  \country{Argentina}
}
\email{maximiliano.arias@fidtech.net}

\author{Alan De Renzis}
\affiliation{%
\institution{Fidtech}
  \streetaddress{Independencia 596}
  \state{Neuquen} 
  \postcode{8300}
  \country{Argentina}}
\email{alan.derenzis@fidtech.net}





\renewcommand{\shortauthors}{M. Garriga et al.}

\begin{abstract}
Blockchain is a decentralized transaction and data management solution, the technological weapon-of-choice behind the success of Bitcoin and other cryptocurrencies. As the number and variety of existing blockchain implementations continues to increase, adopters should focus on selecting the best one to support their decentralized applications (dApps), rather than developing new ones from scratch. In this paper we present a framework to aid software architects, developers, tool selectors and decision makers to adopt the right blockchain technology for their problem at hand. The framework exposes the correlation between technological decisions and architectural features, capturing the knowledge from existing industrial products, technical forums/blogs, experts' feedback and academic literature; plus our own experience using and developing blockchain-based applications. We validate our framework by applying it to dissect the most outstanding  blockchain platforms, i.e., the ones behind the top 10 cryptocurrencies apart from Bitcoin. Then, we show how we applied it to a real-world case study in the insurtech domain. 
\end{abstract}

%
%



\maketitle

\section{Introduction}
\label{sec:intro}

Blockchain is a decentralized transaction and data management solution, well-known for being the technology behind the success of Bitcoin cryptocurrency~\cite{nakamoto2008bitcoin}.
Its main goal is to create a decentralized environment where no third
party is in control of the transactions and data~\cite{huumo2016current}.
This technology is now mainstream because it solves problems in a
way people could not before, generating a business
value-add that will reach about \$176 billion
by 2025 and \$3.1 trillion by 2030\footnote{\url{https://www.gartner.com/doc/3627117/forecast-blockchain-business-value-worldwide}}. The prime reason behind this expansion is the already widespread adoption of blockchain in financial transactions and cross-border
payments~\cite{hileman2017global}. 

Even though the most popular blockchain implementation is Bitcoin,
a myriad others are currently running or still in development. Different
implementations vary in many ways such as their purpose, governance and efficiency, among others. 

However, no single blockchain by itself can meet the requirements
for all usage scenarios, e.g., those that require real-time processing.
When building blockchain-based applications, we need to systematically
consider the key technological features and configurations, and assess their impact on quality attributes for the overall systems~\cite{xu2017taxonomy}.
Moreover, to determine which blockchain implementation should be leveraged
(or even if a new one is needed) for a given application, it is crucial
to be familiar with the differences among them~\cite{Hintzman17scte}.

In this paper we propose a framework to help software architects, developers, tool selectors and decision makers to adopt the right blockchain\footnote{For the sake of simplicity we will use the word blockchain to refer
to any distributed ledger implementation, except when explicitly clarified.} technology for their problem at hand. Even though new blockchains increased exponentially up to 2017, nowadays it makes no sense
to ``reinvent the wheel'' by building a custom blockchain from scratch
every time; but rather to leverage, and probably combine existing, battle-hardened
solutions to support new applications. For crafting our framework, we surveyed the knowledge
from existing industrial products, technical forums/blogs, academic
literature and our own experience using and developing blockchain-based applications.


Afterwards, we applied the framework to analyze and score the most outstanding
solutions in the real-world market \textemdash{} i.e., the technology
behind the top 11 cryptocurrencies\footnote{According to their market cap. See: \url{https://coinmarketcap.com}}, giving an overview of the current ecosystem of mainstream blockchain solutions. We then show how such an assessment
framework can be applied through a real-world case study in the insurtech
domain. 


The rest of the paper is organized as follows. Section~\ref{sec:background}
provides an overview of key concepts in blockchain and cryptocurrencies, as well as related work.
Section~\ref{sec:framework} details our framework for assessment of blockchain technologies. 
Section~\ref{sec:assessment} assesses the top blockchain solutions by means of our framework and then applies it to a real-world case study.
Finally, Section~\ref{sec:conclusion} concludes the paper.

\section{Background}
\label{sec:background}

A \emph{blockchain} is a distributed ledger, in the form of a totally ordered, back-linked list of blocks~\cite{nakamoto2008bitcoin}.
Each block contains transactions that are hashed into a binary hash
tree (also called a \emph{merkle tree}), with the top (root) of the tree stored alongside the transactions. Each block also contains the previous block's hash, thus
guaranteeing integrity and determinism --- i.e., any node replaying all blocks starting from the first one (genesis block) should end up with the same state as every other node~\cite{buterin2014next}. This forbids to call external APIs whose responses may change over time\footnote{\url{http://www.truthcoin.info/blog/contracts-oracles-sidechains/}}. Blockchain distribution is coupled with
trust creation and a consensus mechanism for determining agreement
on the next block to add. \emph{Cryptocurrencies} have emerged as the first generation of blockchain-based applications, as digital currencies that
are based on cryptography techniques and peer-to-peer networks. The
first and most popular example is Bitcoin~\cite{nakamoto2008bitcoin,alharby2017blockchain}.

One of the fundamental disruptions that blockchain technology is causing
is the redefinition of digital \textit{trust}, which manifests itself in a fully distributed way without anyone having to trust any single member of the network. The only trust required is that, on average, the participants of the network are not colluding
against the others in a coordinated manner~\cite{xu2017taxonomy}.

\textit{Transactions }are data packages that store information --- e.g., monetary
value for cryptocurrencies, or results of function calls for other decentralized applications (dApps). The integrity of a transaction is checked by algorithmic rules and cryptographic techniques. A transaction, signed by its initiator, is sent to a node connected to the blockchain network, which validates the transaction and propagates it to other nodes in the network. These also validate and propagate the transaction to their peers, until it reaches all nodes in the network~\cite{xu2017taxonomy}. Transaction processing involves a \textit{transaction fee}, given the cost imposed to the network, and as an incentives for nodes to stay honest~\cite{nakamoto2008bitcoin}. 

Transactions are grouped in blocks that are appended to the existing chain, a process known as \textit{mining}. The network aims to reach a \emph{consensus} about the next block to be included into the blockchain by means of a \textit{Consensus Protocol}. Their
features include assuring decentralized governance, quorum structure, authentication, integrity, non-repudiation, byzantine fault tolerance and performance~\cite{mattila2016blockchain}. The de-facto consensus protocol is Proof-of-Work~\cite{back2002hashcash} (PoW, the one behind Bitcoin and Ethereum), which imposes miners to compute a hash function that should be efficiently verifiable, but parameterisably expensive to compute. Given the ludicrous energy consumption of PoW\footnote{\url{https://digiconomist.net/bitcoin-energy-consumption}}, other blockchains opted for greener but rather centralized options such as Proof-of-Stake~\cite{kiayias2017ouroboros} (PoS), where miner are limited to a percentage of transactions that is reflective of his or her ownership stake of the token, and lately Delegated Proof of Stake (DPoS). Finally, several blockchains provided their own hybrid or ad-hoc protocols~\cite{sankar2017survey}.


The first generation of blockchains (e.g., Bitcoin) provided very limited capability to support programmable transactions, apart from value transfer from one account to another. The second generation (e.g., Ethereum~\cite{buterin2014next}) aims to provide a general-purpose programmable
infrastructure, whose programs are known as \emph{smart contracts}~\cite{szabo1997formalizing,xu2017taxonomy}. Originally, smart contracts were defined as the digital equivalent of a paper contract: an agreement between parties with a set of promises that are legally enforceable~\cite{szabo1997formalizing}. Nowadays, any general purpose computation that takes place on a blockchain or distributed ledger is considered a smart contract. 

\section{Related Work}
\label{sec:related}

Yli-huumo et al.~\cite{huumo2016current} identified that a majority
of the blockchain-related papers focused on certain technical challenges:
throughput, latency, size and bandwidth, security, wasted resources,
usability, versioning, hard forks, and multiple chains~\cite{swan2015blockchain}. In addition, they identified privacy, smart contracts, new cryptocurrencies, botnets, consensus protocols and trustworthiness. As future research
directions, authors highlight scalability issues; other uses beyond cryptocurrency systems (i.e., dApps) and effectiveness
of the proposed solutions --. the latter being one of the contributions of our paper. In a similar direction, Alharby and Van Morsel~\cite{alharby2017blockchain} present a systematic mapping
study of smart contracts. They identified four groups according
to the challenges tackled: codifying, security, privacy and performance.

Scriber~\cite{scriber2018framework} performed a literature review and evaluated 23 blockchain implementation projects. This evaluation revealed 10 architectural or blockchain characteristics that can help determine whether blockchain is appropriate for
a given problem, namely immutability, transparency, trust, identity, distribution, transactions, historical record, ecosystem and inefficiency. However, from the 23 analyzed projects, only four reached an advanced stage while the others failed or were abandoned. The paper does not provide insights regarding which of the most popular blockchains would be suitable for a given problem.

Xu et al.~\cite{xu2017taxonomy}, present a taxonomy of blockchain systems and their architectural characteristics, to assist with the design and assessment of their impact on software architectures. 
First, authors identify fundamental properties of blockchain networks, namely: immutable
data, non-repudiation, data integrity, transparency, equal rights (of participants), and trust mechanisms. Other properties include data privacy, scalability, cost and performance. Afterwards, they distilled architectural design issues, which impact on such properties: decentralization, support for client storage and computation, scope (public/consortium-community/private),
data structure, Consensus protocol and new side-chains. Then, they propose a series of decisions while designing a blockchain-based systems, that may affect those drivers. However, the decisions are intended for the development of a new blockchain rather than evaluating the adoption of an existing one.

\section{A Comparative Framework for blockchain implementations}
\label{sec:framework}

First consideration for a blockchain implementation is its \textit{purpose}. This seems obvious, but is truly overlooked by many
architects and developers. Most existing blockchains are specialized
for cryptocurrencies, and  might not be a good fit for other applications whose intent is different~\cite{Hintzman17scte}. Purposes\footnote{See \url{https://goo.gl/rDAfkK}} are categorized as: 

\begin{itemize}
    \item \textbf{Currencies} used for transactions or as a store of value, e.g., Bitcoin, Litecoin\footnote{\url{https://litecoin.org/}}, Tether~\cite{tether18white}.
    \item \textbf{Exchanges and Interoperability} designed to enable communication among different blockchains, e.g., Binance Coin\footnote{\url{https://www.binance.com/en}}, 0x\footnote{\url{https://0xproject.com/}}.
    \item \textbf{Data and Cloud Services} used to interact with data management or cloud service platforms, e.g., Golem\footnote{\url{https://golem.network/}}.
    \item \textbf{dApps Platforms}:  used as part of a smart contract network or dApps platform, e..g, Ethereum, Cardano~\cite{kiayias2017ouroboros}, EOS~\cite{eos18white}.
    \item \textbf{Gaming, Media, and Social} used for gaming, online content and social media, e.g., Steem\footnote{\url{https://steem.io/}}, Tron~\cite{tron18white}.
    \item \textbf{Privacy}, with built-in features to facilitate anonymous
or untraceable transactions online, e.g., Monero\footnote{\url{https://getmonero.org/}}, Zcash\footnote{\url{https://z.cash/}}.
    \item \textbf{FinTech} for financial services and technologies, e.g., Ripple~\cite{schwartz2014ripple}, Stellar~\cite{mazieres2015stellar}.
    \item \textbf{Business/Enterprise} helps businesses improve efficiency,
transparency, and security, e.g., Waltonchain\footnote{\url{https://www.waltonchain.org/}}.
    \item \textbf{Others}, for example those related to prediction markets, oracles, IoT or AI projects, e.g., IOTA~\cite{popov17tangle}.
\end{itemize}

In the following sections, we define the architectural features to guide this analysis. Alongside, we define technological  decisions and the possible values that they may assume, which finally impact on the features (see Table~\ref{tab:arch-tech} for a summary).

\subsection{Cost}

Altough adopting a blockchain is theoretically free,  at
least three aspects impose a cost for using the network:
a variable cost for running transactions, composed by
the \emph{transaction fee}~\cite{buterin2014next} and \emph{incentives} for processing transactions; and a minimal fixed cost to deploy applications (in the form of smart contracts).

The default approach is to have purely voluntary fees with dynamic minimums~\cite{nakamoto2008bitcoin}. However, this approach can become prohibitively expensive when the network is congested: for example, transaction fees in Bitcoin have raised up to 40 USD during peaks of workload.

On the other end, implementations such as Ripple and Tron foster minimal transaction fees to prevent malicious users to perform DDoS attacks for free. Sitting in the middle, a widely used approach is to define a cost per instruction and then calculate the overall cost of the transaction (as in Ethereum), with a maximum limit in order to avoid infinite loops. Yet another approach is to impose a non-monetary fee for running transactions. For example, in IOTA, every node sending a transaction is required to validate two other transactions, which assures enough processing power.


Possible values for the \emph{transaction fee }are: Minimal (only
to avoid DDoS attacks), per transaction and per instruction. Possible values for the \emph{incentive} are: Big (bitcoin-alike) and small (equivalent to a fee). 

\subsection{Consistency}

Different strategies have been used to confirm that a transaction
is securely appended to the blockchain \textendash{} that is, to ensure
strong \emph{consistency}: wait for a certain
number of blocks (e.g., 6 for Bitcoin, 12 for Ethereum) to have been
generated after the transaction is strong consistent into the blockchain\cite{xu2017taxonomy}; add a checkpoint to the blockchain, so that all the participants will accept the transactions up to the checkpoint as valid and irreversible. Other implementations of the distributed ledger (e.g., a DAG) can
drastically reduce the time to confirmation as they do not rely on
blocks with multiple transactions, but in transactions that are propagated independently in a matter of seconds.

Thus, consistency is a function of the \emph{time to confirmation}
(i.e., the number of blocks after which one can consider a transaction
securely appended to the blockchain), which in turn depends on the
\emph{block production rate} (BPR, the amount of time required to
mine a block), configured for each implementation at design
time.

Possible values for \emph{time to confirmation} are: seconds, minutes
and hours. Possible values for \emph{block production rate} are: 10 minutes or
more, 1 to 10 minutes and seconds.

\subsection{Functionality and Extensibility}

Bitcoin's main intent was to become a decentralized  cryptocurrency~\cite{nakamoto2008bitcoin}. Rapidly, the idea of applying it to other concepts and decentralized application (dApps) emerged, e.g., for name registration and tokens for corporate use~\cite{buterin2014next}, being more flexible and extensible through smart contracts. Some implementations support turing-complete smart contracts using ad-hoc languages (Solidity in Ethereum, Plutus in Cardano), while others support traditional languages (such as C++ in EOS) that are then compiled/transpiled to a bytecode. However, the latter are not fully supported yet in any of the existing blockchain implementation.

A latter point is \emph{interchain communication}, allowing multiple parallel blockchains to interoperate retaining their security properties~\cite{kwon2016cosmos}. Some blockchain implementations provide native support for interchain communication (e.g., EOS), while certain frameworks allow it on top of existing implementations (e.g., Cosmos~\cite{kwon2016cosmos}).

Possible values for \emph{smart contracts }are: Yes (specifying the language(s)), No, and Very Limited. Possible values for \emph{Interchain communication} are: Yes, No.

\subsection{Performance and Scalability}

Decoupling \emph{performance} (latency and throughput)
and \textit{scalability} (with the number of nodes and clients in
the system) is not entirely possible~\cite{vukolic2015quest}, thus
we will group them together for analysis. Those became the bottleneck
for the most popular blockchain implementations, such as Bitcoin (consensus latency of about an hour) and
Ethereum. Additionally, PoW-based networks use a lot of power, equivalent to a small country such as Austria~\cite{vukolic2015quest}. Thus performance and scalability of permisionless generic blockchains is
limited by their design decisions~\cite{vukolic2017rethinking}, namely sequential execution of transactions and hard-coded consensus
protocol.

Scalability, in turn, refers to the ability to maintain performance
indicators when serving more users and transactions, limited by: (i) the size of the data on blockchain, (ii) the transaction processing rate, and (iii) the latency of data transmission. Roughly speaking,
PoW offers good scalability with poor performance, whereas other protocols offer good performance 
for small numbers of replicas, with limited scalability. Given seemingly inherent tradeoffs between the number of nodes
and performance, it is not clear today what the optimal blockchain solution is, for the majority of use cases in which the number of
nodes ranges from a few tens to a few thousands.

Conclusively, performance and scalability are affected by the \emph{transactions per second} (TPS), \emph{block production rate, consensus protocol} and certain \emph{technological choices}.

Possible values for\emph{transactions per second} are: less than 100, between 100 and 1000; and 1000 or more. Possible values for the \emph{Consensus Protocol} are: Proof-of-Work (PoW), Proof-of-Stake (PoS), Distributed Proof-of-Stake (DPoS), and other (specify). Possible values for technological choices are: Merkle/Patricia trees, Segregated Witness (segwit), data sharding, parallel execution of
transactions, GHOST (Greedy Heaviest Observed Sub-Tree), Lightning Network, and other (specify).

\subsection{Security}
 
One of the main features of blockchain is that its public ledger cannot be modified or deleted after the
data has been approved by all nodes, providing data integrity and security characteristics~\cite{huumo2016current}.
Security issues mean bugs or vulnerabilities that an adversary might
utilize to launch an attack. Currently, the most secure implementations are PoW-based. Even though, they have a possibility of a 51\% attack, where a single entity would have full control of the majority of the network's mining hash-rate and would be able to manipulate it.

Alternative consensus protocols such as PoS and DPoS may provide better
performance and/or scalability, but they imply a tradeoff w.r.t. security:
most tolerate up to 1/3 (33\%) of malicious nodes. Other algorithms
implementing Byzantine Fault Tolerance (BFT, e.g., the Ripple Consensus
Protocol) may improve security up to 2/3 malicious nodes, but they
impose additional restrictions such as requiring nodes to know each
other.


Security is thus affected by the following technological decisions:
\emph{Fault tolerance} (possible values are 2/3 attack, 1/2 attack,
1/3 attack, and other); \emph{ledger implementation} (possible values
are Blockchain, DAG, and other); and \emph{consensus protocol}.

\subsection{Decentralization}

Theoretically, blockchain does not rely on any centralized node or authority, allowing data to be recorded, stored and updated in a distributed fashion. However, some blockchains introduce certain degree of centralization. In case of public, permissionless blockchain, no centralized authority or party has more power than the rest (Bitcoin), and everyone has the right to validate a transaction~\cite{vukolic2015quest}.

In the case of consortium, permissioned blockchain, only few nodes are given certain privileges over validation (PoS- and DPoS-based
ones such as EOS). A fully private blockchain has a centralized structure with the power to take decisions and control the validation process (e.g., Ripple and the Ripple Consensus Protocol). Permissioned blockchains are faster, more energy efficient
and easily implementable compared to permissionless blockchains, but introduce certain degree of centralization. Thus, the decentralization degree is constrained by the \emph{consensus
protocol}, and the \emph{ledger implementation}.

\begin{table}

\caption{Correlation between Architectural Features and Technological Decisions\label{tab:arch-tech}}

\begin{tabular}{>{\centering}p{1.3cm}|>{\centering}p{0.7cm}>{\centering}p{0.8cm}>{\centering}p{0.8cm}>{\centering}p{0.8cm}>{\centering}p{1cm}>{\centering}p{0.8cm}}
\hline 
\multirow{2}{1.5cm}{{\footnotesize{}Technological Decision}} & \multicolumn{6}{c}{{\footnotesize{}Architectural Feature}}\tabularnewline
\cline{2-7} 
 & {\footnotesize{}Cost} & {\footnotesize{}Consis\-tency} & {\footnotesize{}Function\-ality} & {\footnotesize{}Perfor\-mance} & {\footnotesize{}Secu\-rity} & {\footnotesize{}Decentra\-lization} \tabularnewline
\hline 
{\footnotesize{}Fees} & {\footnotesize{}x} &  &  &  &  &   \tabularnewline
\hline 
{\footnotesize{}Incentive} & {\footnotesize{}x} &  &  &  &  &  \tabularnewline
\hline 
{\footnotesize{}Confirmation Time} &  & {\footnotesize{}x} &  &  &  &  \tabularnewline
\hline 
{\footnotesize{}Block Production Rate} &  & {\footnotesize{}x} &  & {\footnotesize{}x} &  &  \tabularnewline
\hline 
{\footnotesize{}Smart Contracts} &  &  & {\footnotesize{}x} &  &  &  \tabularnewline
\hline 
{\footnotesize{}Interchain} &  &  & {\footnotesize{}x} &  &  &  \tabularnewline
\hline 
{\footnotesize{}Consensus} &  &  &  & {\footnotesize{}x} &  & {\footnotesize{}x}\tabularnewline
\hline 
{\footnotesize{}Technology} &  &  &  & {\footnotesize{}x} &  & \tabularnewline
\hline 
{\footnotesize{}Fault Tolerance} &  &  &  &  & {\footnotesize{}x} &  \tabularnewline
\hline 
{\footnotesize{}Ledger} &  &  &  &  & {\footnotesize{}x} & {\footnotesize{}x} \tabularnewline
\hline 
{\footnotesize{}TPS} &  &  &  & {\footnotesize{}x} &  &  \tabularnewline
\hline 
\end{tabular}

\end{table}

\section{Assessment of Blockchain Solutions}
\label{sec:assessment}

The initial list of features and decisions extracted from the literature was a subset of the ones in Table~\ref{tab:arch-tech}. Those were delivered to a group of three experts, comprising researchers and industry practitioners. They evaluated the list and suggested to add, remove, group, or decompose concepts, and pointed out correlations, based on their own expertise. 

Then, we proceeded to extract the technological features lying under the most popular blockchain implementations, according to their market cap\footnote{source: \url{http://coinmarketcap.com}, July 2018}. Even though we acknowledge other possible ways to measure popularity
\textemdash{} e.g., number of wallets, number of exchanges, transaction volume, etc. \textemdash{} they usually converge to a similar ranking at a given point in time~\cite{hileman2017global}. Popularity is not affected by the technical decisions behind the implementation, but may impact certain features of the blockchain such as cost, time to confirmation and security. Additionally, the popularity of the underlying blockchain may be the key enabler for the success of a dApp, as demonstrated by the myriad of dApps in Ethereum\footnote{\url{https://github.com/avadhootkulkarni/UltimateICOCalendar}} and the limited offer in others.
All in all, the technological analysis of the different blockchains is summarized in Table~\ref{tab:qualitative-analysis}.

\begin{sidewaystable*}
\caption{Qualitative assessment of the most popular blockchain alternatives
according to the technological characteristics\label{tab:qualitative-analysis}}

\tabcolsep=4pt

\begin{tabular}{|>{\raggedright}p{1.8cm}|>{\raggedright}p{1.6cm}|>{\raggedright}p{1.6cm}|>{\raggedright}p{1.6cm}|>{\raggedright}p{1.6cm}|>{\raggedright}p{1.6cm}|>{\raggedright}p{1.6cm}|>{\raggedright}p{1.6cm}|>{\raggedright}p{1.5cm}|>{\raggedright}p{1.6cm}|>{\raggedright}p{1.6cm}|>{\raggedright}p{1.6cm}|}
\hline 
 & \textbf{\footnotesize{}Bitcoin}\\

\textbf{\footnotesize{}(BTC)} & \textbf{\footnotesize{}Ethereum}\\
\textbf{\footnotesize{} (ETH)} & \textbf{\footnotesize{}Ripple}\\
\textbf{\footnotesize{} (XRP)} & \textbf{\footnotesize{}Bitcoin Cash}\\
\textbf{\footnotesize{} (BCH)} & \textbf{\footnotesize{}EOS }\\
\textbf{\footnotesize{}(EOS)} & \textbf{\footnotesize{}Litecoin}\\
\textbf{\footnotesize{} (LTC)} & \textbf{\footnotesize{}Stellar}\\
\textbf{\footnotesize{} (XLM)} & \textbf{\footnotesize{}Cardano}\\
\textbf{\footnotesize{} (ADA)} & \textbf{\footnotesize{}Iota}\\
\textbf{\footnotesize{} (MIOTA)} & \textbf{\footnotesize{}Tether}\\
\textbf{\footnotesize{} (USDT)} & \textbf{\footnotesize{}Tron}\\
\textbf{\footnotesize{} (TRX)}\tabularnewline
\hline 
\textbf{\footnotesize{}Purpose} & {\footnotesize{}Currency} & {\footnotesize{}Platform} & {\footnotesize{}Fintech} & {\footnotesize{}Currency} & {\footnotesize{}Platform} & {\footnotesize{}Currency} & {\footnotesize{}Fintech} & {\footnotesize{}Platform} & {\footnotesize{}Other (IoT)} & {\footnotesize{}Currency} & {\footnotesize{}Media/Social}\tabularnewline
\hline 
\textbf{\footnotesize{}Fees} & {\footnotesize{}Per transaction} & {\footnotesize{}Per Instruction (configurable)} & {\footnotesize{}Minimal} & {\footnotesize{}Per Transaction} & {\footnotesize{}Per instruction} & {\footnotesize{}Per transaction} & {\footnotesize{}Per operation} & {\footnotesize{}Per transac\-}\\
{\footnotesize{}tion size} & {\footnotesize{}Approve 2 transactions} & {\footnotesize{}20 USD} & {\footnotesize{}Minimal}\tabularnewline
\hline 
\textbf{\footnotesize{}Incentive} & {\footnotesize{}12.5 BTC/hash} & {\footnotesize{}Configurable} & {\footnotesize{}N/A} & {\footnotesize{}12.5 BCH/hash} & {\footnotesize{}Configurable (by BPs)} & {\footnotesize{}6.5 LTC} & {\footnotesize{}N/A} & {\footnotesize{}Availability based} & {\footnotesize{}N/A} & {\footnotesize{}No} & {\footnotesize{}Configurable (by SRs)}\tabularnewline
\hline 
\textbf{\footnotesize{}Block~ Prod\-uction~rate} & {\footnotesize{}10 min} & {\footnotesize{}10-19 sec} & {\footnotesize{}Not specified} & {\footnotesize{}10 min} & {\footnotesize{}0.5 sec} & {\footnotesize{}2.5 min} & {\footnotesize{}5 seconds} & {\footnotesize{}20 sec configurable} & {\footnotesize{}N/A} & {\footnotesize{}10 min} & {\footnotesize{}15 sec}\tabularnewline
\hline 
\textbf{\footnotesize{}Confirma\-tion~time} & {\footnotesize{}60 min (6 blocks)} & {\footnotesize{}1 min} & {\footnotesize{}seconds} & {\footnotesize{}60 min} & {\footnotesize{}1 second} & {\footnotesize{}20 min (6 blocks)} & {\footnotesize{}30 seconds} & {\footnotesize{}1 min} & {\footnotesize{}2 min (scalable)} & {\footnotesize{}60 min (6 blocks)} & {\footnotesize{}1 min}\tabularnewline
\hline 
\textbf{\footnotesize{}TPS} & {\footnotesize{}7} & {\footnotesize{}15} & {\footnotesize{}1500} & {\footnotesize{}23} & {\footnotesize{}1000+} & {\footnotesize{}50} & {\footnotesize{}1000+} & {\footnotesize{}250} & {\footnotesize{}1000+} & {\footnotesize{}7} & {\footnotesize{}1000+}\tabularnewline
\hline 
\textbf{\footnotesize{}Smart contracts} & {\footnotesize{}Very Limited} & {\footnotesize{}Yes} & {\footnotesize{}Very Limited} & {\footnotesize{}Very Limited} & {\footnotesize{}Yes} & {\footnotesize{}Very Limited} & {\footnotesize{}Very Limited} & {\footnotesize{}Yes} & {\footnotesize{}No} & {\footnotesize{}No} & {\footnotesize{}Yes}\tabularnewline
\hline 
\textbf{\footnotesize{}dApps} & {\footnotesize{}No} & {\footnotesize{}Yes} & {\footnotesize{}No} & {\footnotesize{}No} & {\footnotesize{}Yes} & {\footnotesize{}No} & {\footnotesize{}No} & {\footnotesize{}Yes} & {\footnotesize{}Yes, IoT level} & {\footnotesize{}Very Limited} & {\footnotesize{}Yes}\tabularnewline
\hline 
\textbf{\footnotesize{}Languages} & {\footnotesize{}C++, Script} & {\footnotesize{}Solidity} & {\footnotesize{}Any} & {\footnotesize{}Script} & {\footnotesize{}Any} & {\footnotesize{}Script} & {\footnotesize{}Any} & {\footnotesize{}Functional} & {\footnotesize{}Any} & {\footnotesize{}Ruby} & {\footnotesize{}Java}\tabularnewline
\hline 
\textbf{\footnotesize{}Interchain} & {\footnotesize{}No} & {\footnotesize{}No} & {\footnotesize{}No} & {\footnotesize{}No} & {\footnotesize{}Yes} & {\footnotesize{}No} & {\footnotesize{}No} & {\footnotesize{}Yes} & {\footnotesize{}Yes} & {\footnotesize{}No} & {\footnotesize{}Yes}\tabularnewline
\hline 
\textbf{\footnotesize{}Consensus Algorithm} & {\footnotesize{}PoW (sha-256)} & {\footnotesize{}PoW (ethash)} & {\footnotesize{}PoC (Ripple CP)} & {\footnotesize{}PoW (sha-256)} & {\footnotesize{}Delegated PoS } & {\footnotesize{}PoW (scrypt)} & {\footnotesize{}Federated BFT~Stellar CP} & {\footnotesize{}PoS (Ourboros)} & {\footnotesize{}Markov Chain (MCMC)} & {\footnotesize{}PoR (Omni)} & {\footnotesize{}PoS (Tendermint)}\tabularnewline
\hline 
\textbf{\footnotesize{}Enhanced Technology} & {\footnotesize{}Merkle Trees, Segwit} & {\footnotesize{}Patricia Trees, Sharding} & {\footnotesize{}\textendash{}} & {\footnotesize{}Merkle Trees} & {\footnotesize{}Merkle Proofs, Segwit} & {\footnotesize{}Merkle Trees, Segwit} & {\footnotesize{}\textendash{}} & {\footnotesize{}Sharding} & {\footnotesize{}Tangle for IoT Scale} & {\footnotesize{}Merkle trees, Segwit} & {\footnotesize{}Graph database}\tabularnewline
\hline 
\textbf{\footnotesize{}Fault Tolerance} & {\footnotesize{}51\% CPU attack} & {\footnotesize{}51\% ether attack} & {\footnotesize{}20\% malus nodes} & {\footnotesize{}51\% CPU attack} & {\footnotesize{}66\% Block Producers} & {\footnotesize{}51\% CPU Attack} & {\footnotesize{}Asymptotic Security} & {\footnotesize{}Asymptotic Security} & {\footnotesize{}51\% CPU attack} & {\footnotesize{}51\% CPU Attack} & {\footnotesize{}33\% Byzantine Failure}\tabularnewline
\hline 
\textbf{\footnotesize{}Ledger} & {\footnotesize{}Blockchain} & {\footnotesize{}Blockchain} & {\footnotesize{}Interledger Protocol} & {\footnotesize{}Blockchain} & {\footnotesize{}Blockchain} & {\footnotesize{}Blockchain} & {\footnotesize{}Blockchain} & {\footnotesize{}Blockchain} & {\footnotesize{}Directed~ Acy\-clic~ Graph} & {\footnotesize{}Blockchain} & {\footnotesize{}Blockchain}\tabularnewline
\hline 
\end{tabular}
\end{sidewaystable*}

Based on the identified architectural and technological aspects, we
conducted a second round of feedback through structured interviews
with the experts, in order to come up with a quantitative assessment
of the top blockchain solutions. They completed a questionnaire\footnote{\url{https://goo.gl/forms/8mE1mdJ55VJRJw2E3}},
assigning scores for each architectural feature to the different blockchain
implementations (from \textit{very low} to \textit{very high}) which
were then fuzzified into numerical scale from 1 to 5. For example,
if an expert considers that \textit{Bitcoin} has a very high \textit{Cost},
then in the questionnaire she marks the corresponding cell as ``very
high''. 

Experts fulfilled the scorecards as described above, also declaring their confidence (fuzzified from \textquotedbl low\textquotedbl{} to \textquotedbl very high\textquotedbl ). Both the confidence values and the scores were then defuzzified using a triangular membership function~\cite{pedrycz94triangular} and combined on a weighted average scheme. The triangular function allows one to map and normalize the
linguistic scale to a given scale, in the range {[}$1,5${]} for the scores and {[}$0,1${]} for the confidence values. 

The result of the process is a normalized weight matrix, which numerically
represent the scores for each feature on each blockchain implementation
as values in the interval {[}$1, 5${]}, as shown in Table~\ref{tab:quantitative-scores}. The information contained on the scorecard allows software engineers, architects and decision makers to assess the different blockchain implementations, being able to select the most suitable one for their problem at hand, as illustrated in the following section.

\begin{table*}
\caption{Quantitative assessment of blockchain solutions according to experts'
feedback regarding architectural decisions.}
\label{tab:quantitative-scores}
\tabcolsep=4pt

{\footnotesize{}}%
\begin{tabular}{>{\raggedright}p{2.2cm}>{\centering}p{0.8cm}>{\centering}p{0.8cm}>{\centering}p{0.8cm}>{\centering}p{0.8cm}>{\centering}p{0.8cm}>{\centering}p{0.8cm}>{\centering}p{0.8cm}>{\centering}p{0.8cm}>{\centering}p{1cm}>{\centering}p{0.8cm}>{\centering}p{0.8cm}}
\toprule 
 & {\footnotesize{}BTC} & {\footnotesize{}ETH} & {\footnotesize{}XRP} & {\footnotesize{}BCH} & {\footnotesize{}EOS} & {\footnotesize{}XLM} & {\footnotesize{}LTC} & {\footnotesize{}ADA} & {\footnotesize{}USDT} & {\footnotesize{}MIOTA} & {\footnotesize{}TRX}\tabularnewline
\midrule
\emph{\footnotesize{}Popularity} & \emph{\footnotesize{}1} & \emph{\footnotesize{}2} & \emph{\footnotesize{}3} & \emph{\footnotesize{}4} & \emph{\footnotesize{}5} & \emph{\footnotesize{}6} & \emph{\footnotesize{}7} & \emph{\footnotesize{}8} & \emph{\footnotesize{}9} & \emph{\footnotesize{}10} & \emph{\footnotesize{}11}\tabularnewline
\midrule
\midrule 
{\footnotesize{}Cost} & {\footnotesize{}1.33} & {\footnotesize{}2} & {\footnotesize{}4.66} & {\footnotesize{}1.66} & {\footnotesize{}5} & {\footnotesize{}4.66} & {\footnotesize{}2.66} & {\footnotesize{}4.33} & {\footnotesize{}5} & {\footnotesize{}5} & {\footnotesize{}5}\tabularnewline
\midrule 
{\footnotesize{}Consis\-tency} & {\footnotesize{}1.33} & {\footnotesize{}2.33} & {\footnotesize{}4.33} & {\footnotesize{}1.33} & {\footnotesize{}5} & {\footnotesize{}4} & {\footnotesize{}2} & {\footnotesize{}3.66} & {\footnotesize{}1} & {\footnotesize{}4.66} & {\footnotesize{}4}\tabularnewline
\midrule 
{\footnotesize{}Function\-ality} & {\footnotesize{}2} & {\footnotesize{}5} & {\footnotesize{}1.33} & {\footnotesize{}2} & {\footnotesize{}5} & {\footnotesize{}1.33} & {\footnotesize{}2} & {\footnotesize{}4.33} & {\footnotesize{}2} & {\footnotesize{}3.66} & {\footnotesize{}5}\tabularnewline
\midrule 
{\footnotesize{}Perfor\-mance} & {\footnotesize{}1.33} & {\footnotesize{}1.66} & {\footnotesize{}4.33} & {\footnotesize{}2} & {\footnotesize{}4.66} & {\footnotesize{}4} & {\footnotesize{}2.33} & {\footnotesize{}3} & {\footnotesize{}1} & {\footnotesize{}5} & {\footnotesize{}4.66}\tabularnewline
\midrule 
{\footnotesize{}Secu\-rity} & {\footnotesize{}4} & {\footnotesize{}4} & {\footnotesize{}2.33} & {\footnotesize{}4} & {\footnotesize{}3.33} & {\footnotesize{}4} & {\footnotesize{}4} & {\footnotesize{}4} & {\footnotesize{}3.33} & {\footnotesize{}3.66} & {\footnotesize{}3.33}\tabularnewline
\midrule 
{\footnotesize{}Decentralization} & {\footnotesize{}5} & {\footnotesize{}3.33} & {\footnotesize{}1} & {\footnotesize{}4.33} & {\footnotesize{}2.66} & {\footnotesize{}2.33} & {\footnotesize{}3.66} & {\footnotesize{}3.33} & {\footnotesize{}1.33} & {\footnotesize{}2.33} & {\footnotesize{}3.33}\tabularnewline
\midrule 
\textbf{\footnotesize{}Total} & {\footnotesize{}14.99} & {\footnotesize{}18.32} & {\footnotesize{}17.98} & {\footnotesize{}15.32} & {\footnotesize{}25.65} & {\footnotesize{}20.32} & {\footnotesize{}16.65} & {\footnotesize{}22.65} & {\footnotesize{}13.66} & {\footnotesize{}24.31} & {\footnotesize{}25.32}\tabularnewline
\midrule 
\end{tabular}{\footnotesize\par}
\end{table*}

\subsection{Open Challenges}

From the analysis of literature, top blockchain implementations, and feedback from experts, we were able to identify some open challenges and concerns regarding the future of the field. Although all blockchain implementations promise to be secure and efficent, most of them fall short in some of these aspects. Particularly, Proof-of-Stake and Distributed Proof-of-Stake blockchains are risky since critical decisions fall on a small group of people or company. Even though, in traditional implementations based on Proof-of-Work, grouping of miners into mining pools are effectively centralizing these networks~\cite{gervais2014bitcoin}. 

In this direction, platforms with the potential to be scalable and energy-efficient will be the weapon-of-choice for dApps development. Other ``legacy'' blockchains such as Bitcoin will remain for big, sporadic transactions and as long-term investment.
    
Blockchain is still an emerging technology, thus not a lot of developers are concerned with the principles of Software Engineering applied to blockchain-based systems. Moreover, the lack of guidelines and standards on how to design software architectures that include smart contracts as part of the system calls for further attention~\cite{destefanis2018smart,xu2017taxonomy}. 

Finally, only a handful of experiences in real-world dApps exist\footnote{\url{https://dappradar.com/}}, and still a lot of controversy on whether an application requires the use of blockchain. The emergence of frameworks like the one presented in this paper may help to overcome such difficulties.

As threats to validity for our approach, we can highlight the following. First, the short number of experts that participated in the analysis. This might result on a possible bias, but avoids the answers to be meaningless because of the lack of experience of surveyed experts. One should also note that the number of experts in the blockchain world is not that big. Another concern is the number of analyzed cryptocurrencies, since among the top ones there is Bitcoin itself and two Bitcoin forks (Litecoin, Bitcoin Cash). Some revolutionary approaches may have not gained momentum yet, reason why we are planning to extend our analysis, covering more implementations.

\subsection{Case study: A trusted images application for the insurtech domain}
\label{sec:case-study}

In this section we illustrate the value of our framework to assess blockchain alternatives for a real-world application. Photofied\footnote{https://photofied.tech} is a mobile application developed by Fidtech, as a solution for worldwide insurance activity. It certifies digital images in the blockchain for fraud prevention, granting reliability of the status of an insurable risk, both at policy emission
and execution stages. 
All images taken with Photofied are certified by means of an ad-hoc protocol, namely Three Way Certification (3WC). 3WC features blockchain, a P2P distributed file system and digital signature to grant the immutability, perdurability and verifiability of the images. 

\begin{figure}
    \centering
    \includegraphics[width=60mm]{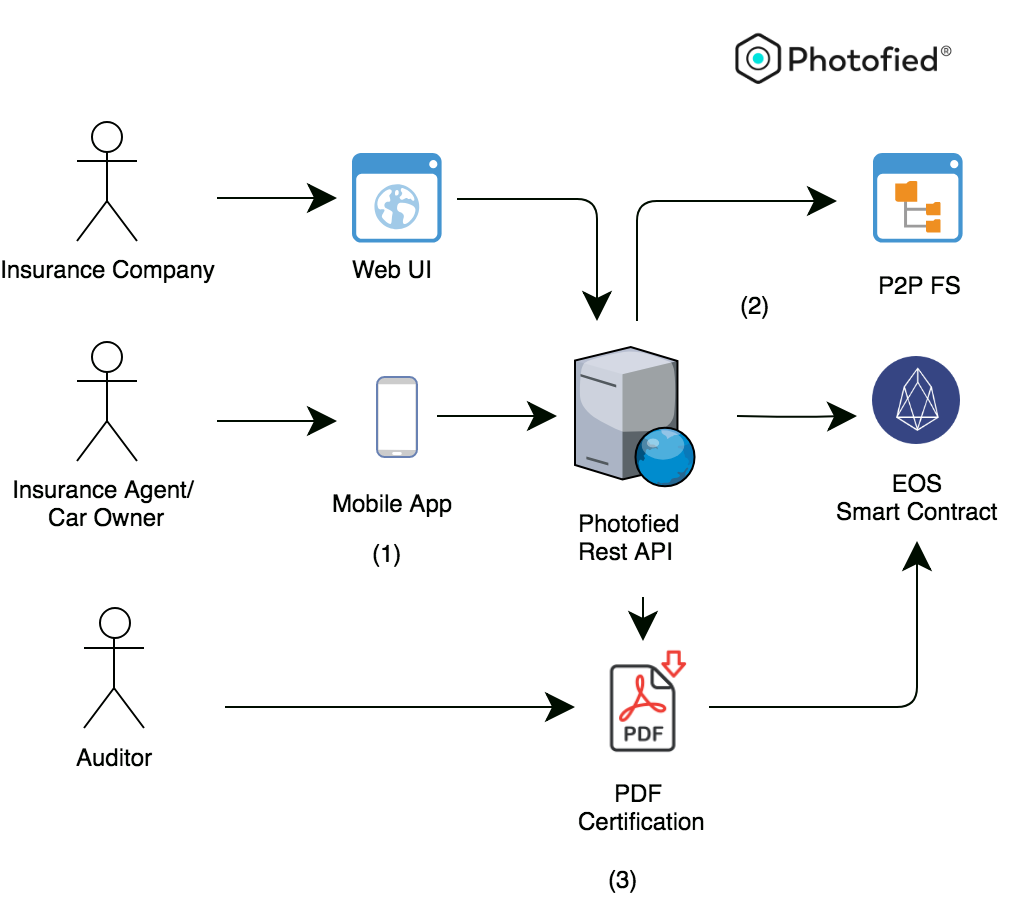}
    \caption{Photofied architecture overview}
    \label{fig:photofiedArch}
\end{figure}

Figure \ref{fig:photofiedArch} depicts the architecture of the application, where insurance agents or car owners use the mobile app (1) to send packages (containing images and metadata to certify) to the Rest API. The server forwards the package to the EOS smart contract and the p2p FS (2). After that, each certification is printed to PDF, allowing offline audit (3). At any time, insurance companies can access the certifications using a Web interface. 

As a first stage, Photofied is used during the policy emission process, certifying the images taken by the insurance agent. The images are later audited only if needed, thus there is no need for fast transaction confirmation. 

End users are neither supposed to know about blockchain or cryptocurrencies and/or own accounts; nor responsible for paying for the service --- thus, Cost should be low to attract insurance companies as potential customers.

Also, as images can be captured either by insurance agents or car owners, the application needs to identify who took the images, when, and where, providing functionality and flexibility.
Each image should be independently certified, which implies a high number of transactions, calling for performance. Additionally, security and consistency are concerns to maximize, granting the trustability of certified images.
Each certification, containing images and metadata (username, GPS coordinates and mobile device's information), has to be auditable by third parties without using Photofied services/servers (i.e., by querying directly the underlying blockchain).


At application's design time, the developers were not aware of all the advantages and drawbacks of each blockchain platform. A first selection, purely based on popularity, led first to a Bitcoin-based implementation and then an Ethereum-based one. The former used a na\"ive (\textit{Data hash} $\rightarrow$ \textit{Timestamp}) structure, given the lack of proper smart contracts support in the Bitcoin blockchain. The latest used a smart contract that stored a mapping from each uploaded piece of information to the account from which it was uploaded along with some metadata.

However, both implementations suffered the drawbacks of the underlying blockchains (See Table~\ref{tab:qualitative-analysis} and Table~\ref{tab:quantitative-scores}), requiring an expensive transaction fee and relying on congested networks, with a prohibitively low number of transactions per second. In parallel, the number of novel blockchain platforms increased exponentially, paving the way for the adoption of a most suitable one. After crafting the comparison framework and fine-tunning the importance for each feature, the most suitable options were EOS and TRON, due to the nonexistent fee, high transactions per second and high reliability. To untie, EOS was selected based on its popularity, as it implies more active developers, nodes, available dApps and supporting community. Also, by that time, the TRON mainnet was not yet online and had no near release date. 

All in all, the current version of Photofied is running using EOS in a collaborate effort with EOS Argentina\footnote{https://www.eosargentina.io/}, one of the top block producers on the EOS blockchain. The smart contract that handles all the needed data is managed by a custom account in charge of time-stamping transactions --- combining block timestamp and server timestamp. This way, end users don't need an EOS account, the application can run on mobile devices as it uses a central server that manages the transactions (ensuring high transactions per second and reliability). Finally, as EOS is a public, decentralized blockchain, each piece of certified data can be audited from any EOS node. 


\section{Conclusions and Future Work}
\label{sec:conclusion}

Nowadays, the number and variety of existing blockchain implementations continues to increase. Adopters should focus on selecting the best one --- rather than developing yet another one from scratch --- to support their decentralized applications (dApps).
In this paper we presented a framework to aid software architects, developers, tool selectors and decision makers to adopt the right blockchain technology for their problem at hand. The framework exposes the correlation between technological decisions (such as consensus protocols and support for smart contracts) and architectural decisions (such as cost and decentralization). For crafting our framework, we surveyed the knowledge from existing industrial products, technical forums/blogs, experts' feedback and academic literature; plus our own experience using and developing blockchain-based applications. 

We have shown the suitability of our framework in two ways. First, we applied it to analyze the most popular blockchain implementations in the real world, according to their market cap. This shed light regarding the current ecosystem of mainstream blockchain solutions. Second, we shown how the framework can be applied by dApps developers through a real-world case study: a trusted images application for the insurtech domain. Developers were able to successfully select a new blockchain and migrate their application based on the insights obtained from the framework.

Our future work comprises fine-tunning the framework by engaging yet more experts from the blockchain world. Afterwards we plan to assess the top 50 implementations to have a complete panorama of the existing solutions, beyond the Bitcoin and Ethereum hype. Finally, we are currently developing a series of questions in the form of a wizard, to guide practitioners in the use of our framework for selecting the most suitable blockchain.

\begin{acks}
 
  The authors would like to thank to Nicolas Arias, Diego Anabalon, Nahuel Vazquez and Claudio Arce from Fidtech, for their helpful insights. This work is partially supported by ANPCyT project PICT-1725-2017 and CONICET.  
  
\end{acks}


\end{document}